\documentclass[aps,prb,twocolumn,superscriptaddress]{revtex4-2}
\usepackage{amsmath}
\usepackage{mathtools} 
\usepackage{amssymb}
\usepackage{amsthm}
\usepackage{bm} 
\usepackage{upgreek} 
\usepackage{xcolor}
\usepackage{graphicx}
\usepackage{epstopdf}
\usepackage{braket}
\usepackage[colorlinks,linkcolor=blue,anchorcolor=blue,citecolor=blue,urlcolor=blue]{hyperref}

\newcommand{\blue}[1]{{\color{blue}#1}}

\newcommand{\up}{\uparrow}
\newcommand{\dn}{\downarrow}

\newcommand{\av}[1]{\left\langle #1 \right\rangle}

\newcommand{\me}{\mathrm{e}}

\begin{document}
\title{Determinant quantum Monte Carlo for the half-filled Hubbard model with nonlocal density-density interactions}

\author{Meng Yao} 
\affiliation{National Laboratory of Solid State Microstructures $\&$ School of Physics, Nanjing University, Nanjing 210093, China}

\author{Da Wang} \email{dawang@nju.edu.cn}
\affiliation{National Laboratory of Solid State Microstructures $\&$ School of Physics, Nanjing University, Nanjing 210093, China}
\affiliation{Collaborative Innovation Center of Advanced Microstructures, Nanjing University, Nanjing 210093, China}

\author{Qiang-Hua Wang} \email{qhwang@nju.edu.cn}
\affiliation{National Laboratory of Solid State Microstructures $\&$ School of Physics, Nanjing University, Nanjing 210093, China}
\affiliation{Collaborative Innovation Center of Advanced Microstructures, Nanjing University, Nanjing 210093, China}

\begin{abstract}
We design a novel formalism of determinant quantum Monte Carlo method for the half-filled Hubbard model with on-site Hubbard interaction $U$ and nearest neighbor density-density interaction $V$ on the square lattice. The formalism is free of sign problem for $|U|\ge 4|V|$, and is achieved by introducing discrete auxiliary fields on the nearest-neighbor bonds alone. Based on this formalism, we study the ground state phase diagram of the model systematically {using the projector algorithm}. Within the sign-problem free parameter space $|U|\ge 4|V|$, we obtain antiferromagnetism for $U\ge 4|V|>0$, charge density wave for $U<-4V$ and $V>0$, s-wave superconductivity for $U<0$ and small negative $V$, and phase separation for $U<0$ and a larger negative $V$. We also obtain the single particle gap and spin excitation spectra, and by comparison to mean field results, as well as available literatures, we discuss the possible phase boundary beyond the sign-problem free region. The unbiased numerically exact results can be taken as benchmarks in future studies. Finally, we discuss possible extension for longer-range interactions in the model.
\end{abstract}

\maketitle

\section{INTRODUCTION}

The Hubbard model \cite{hubbard,Gutzwiller1963} is a standard model for strongly correlated electron systems and has been studied extensively in the past few decades, not only because of its fundamental theoretical significance \cite{Arovas2022} but also because of its close relation to real materials, such as high temperature superconductors \cite{Anderson1987, Lee2006} and cold atoms \cite{Bloch2008}.
In the simplest form, the interaction in the model contains the local Hubbard interaction $U$ alone, but the extension to near-neighbor interactions is straightforward. Such interactions are repulsive if they are derived from pure Coulomb interaction, but could also be effectively attractive if they are caused by electron-phonon coupling, or could be tuned to be attractive in the context of cold atoms. Interestingly, an attractive nearest-neighbor interaction $V$ seems to be present and plays an important role in one-dimensional cuprate chains \cite{Chen2021}. Nonlocal Coulomb interactions are also argued to be important in graphene \cite{nonlocal1} because of poor screening, and in magic-angle twisted bilayer graphene \cite{Cao2018} because of the peculiar three-lobe Wannier functions for the low energy electron degrees of freedom \cite{Koshino2018,Kang2018,Po2018}. These progresses renew the interest in the extended Hubbard model with the on-site Hubbard $U$ and density-density interactions $V$ on neighboring bonds \cite{rc1, rc2, rc3, rc5, rc6, rc7, rc8, rc9, rc10, rc11, rc12, rc13, rc14, rc15}.

Although the extended Hubbard model has been widely studied, some of its properties are still unclear. For repulsive $U$ and $V$, the phase transition between antiferromagnetism (AFM) and charge density wave (CDW) is predicted at $U=4V$ in mean field theory (MFT) \cite{Dagotto} and in the strong coupling limit \cite{strong_coupling, Hirsch1984}, but some quantum cluster methods \cite{vcpt, dca1, dca2, dca3} and dual boson approach \cite{dba} locate the phase boundary at $U<4V$. For attractive $V$, the existence of phase separation (PS) \cite{1d2} and its relation to the superconductivity (SC) are still under debate \cite{ps_rpa, Dagotto, 2d3, ps_db}. Actually, to the best of our knowledge, no large-scale exact numerical simulation has been performed systematically for the ground state phase diagram of the extended Hubbard model with both $U$ and $V$.

The quantum Monte Carlo (QMC) is {a powerful tool} for the study of correlated systems, because it is in principle unbiased and statistically exact. Even the ground state can be accessed by the projector determinant QMC (DQMC) \cite{dqmc}. Unfortunately, depending on the specific model, the QMC method often suffers from the notorious negative sign problem, the resolution of which is one of the most difficult challenges in both physics and mathematics communities \cite{Troyer2005}. Nonetheless, it is known that some symmetries can protect the system from the sign problem, opening up a window to study particular models with appropriate symmetries exactly by QMC. In DQMC, the sign problem is closely related to the Hubbard-Stratonovich (HS) transformation of the interactions. For a given HS configuration, if there is a particle-hole (PH) \cite{Hirsch1985} or time-reversal (TR) \cite{Wu2005} symmetry, or Majorana reflection/Kramers symmetry\cite{Wei2016,Li2016}, the sign problem can be avoided exactly. For example, the attractive Hubbard $U$-term can be decoupled with discrete HS transformation \cite{Hirsch1983} in the charge channel
that spin up and down electrons see the same HS fields and contribute the same real determinant, so that the product of the determinants are positive definite. For repulsive $U$, the auxiliary field becomes pure imaginary. For the half-filled model on a bipartite lattice, however, the sign problem can be proved to vanish by the partial PH transformation for only spin down electrons, after which the spin-up and spin-down determinants are complex conjugation to each other. The product of the determinants are again positive definite. {Moreover, this HS transformation maintains the spin $SU(2)$ symmetry and thus is efficient for the study of spin dynamics \cite{assaad1999}.}

We are interested in the DQMC for the half-filled Hubbard model with both $U$ and $V$. A direct decoupling of the nearest-neighbor density-density interaction $V$ breaks both TR and PH symmetries, causing the sign problem immediately \cite{2d1}.
This problem is partially solved in Ref.~\cite{2d2}, where a HS decoupling in the hopping-channel is employed to treat both $U$ and $V$ terms together. This approach is free of sign problem for $U<-8|V|$ on the square lattice, hence works for negative $U$ only. Further progress is achieved in Refs.~\cite{2d4,2d5}, where a continuous HS decoupling in the charge channel is applied for a bipartite lattice and works without sign problem for $|U|>z|V|$, where $z$ is the coordination number. Later, an improved scheme is proposed in Ref.~\cite{2d6}, where discrete HS decouplings are applied on both sites and nearest-neighbor bonds. The accessible parameter space is also limited by $|U|>z|V|$, but the sampling over discrete HS fields becomes easier and more efficient.

In this work we propose a further improved decoupling scheme for the half-filled bipartite Hubbard model with both $U$ and $V$. It involves discrete HS fields on the nearest-neighbor bonds {\em alone}. We then apply the novel scheme to study systematically the ground state phase diagram of the square lattice. Within the sign-problem free parameter region $|U|\ge 4|V|$, we obtain AFM for $U\ge 4|V|>0$, CDW for $U<-4V$ and $V>0$, s-wave SC for $U<0$ and small negative $V$, and PS for $U<0$ and larger negative $V$. We also obtain the single particle gap and spin excitation spectra, and by comparing with MFT as well as available literatures, we discuss the possible phase boundaries beyond the sign-problem free region. The unbiased numerically exact results can be taken as benchmarks in future studies.

The paper is organized as follows. In Sec.~\ref{section2}, we introduce our HS decoupling scheme and technical details for the measurements in our DQMC. In Sec.~\ref{section3}, we perform systematic simulations for the ground state phase diagram using DQMC with the novel decoupling scheme, and discuss the results in comparison to MFT and previous results in the literature. Sec.~\ref{section4} contains the summary and perspective of this work.

\section{MODEL AND METHODS}\label{section2}
We consider the half-filled extended Hubbard model on the square lattice described by the Hamiltonian
\begin{align} \label{eq:model}
	H=&-t\sum_{\langle ij\rangle \sigma}(c_{i \sigma}^{\dagger}c_{j\sigma}+h.c.)+U\sum_{i}(n_{i \uparrow}-\frac{1}{2})(n_{i \downarrow}-\frac{1}{2})\nonumber\\
	&+V\sum_{\langle ij\rangle} (n_i-1)(n_j-1),
\end{align}
where $c_{i \sigma}$ is the electron annihilation operator at site $i$ and with spin $\sigma$, $n_i = n_{i\uparrow} + n_{i\downarrow}$ is the total local density, and $\langle ij\rangle$ denotes the nearest neighbor bond. We use $t=1$ as the unit of energy henceforth.

We limit ourselves to the ground state at zero temperature, which allows ordered states breaking even continuous symmetries. We use the zero-temperature (projector) DQMC to calculate the expectation value of an operator $\hat{O}$ in the ground state $\ket{\Psi_0}$,
\begin{equation}\label{eq:expection}
	\bra{\Psi_0}\hat{O}\ket{\Psi_0}=\lim_{\Theta\rightarrow\infty}\frac{\bra{\Psi_T}\text{e}^{-\Theta H/2} \hat{O} \text{e}^{-\Theta H/2}\ket{\Psi_T}}{\bra{\Psi_T}\text{e}^{-\Theta H}\ket{\Psi_T}},
\end{equation}
where $\ket{\Psi_T}$ is the trial wave function, which is required to be non-orthogonal to the ground state, $\bra{\Psi_0} \Psi_T\rangle \neq 0$, and $\Theta/2$ is the projection time.
The projector is Trotter-Suzuki decomposed as,
\begin{equation}
	\text{e}^{-\Theta H}=[\text{e}^{-\frac12\Delta \tau H_t}\text{e}^{-\Delta\tau H_{I}}\text{e}^{-\frac12\Delta \tau H_t}]^M+\mathcal{O}(\Delta \tau^3 )
\end{equation}
where $\Theta=\Delta \tau M$, and $H_{t}$ and $H_{I}$ denote the kinetic and interaction terms, respectively. In our simulations, we empirically set $\Theta=2L\sqrt{4/|U|}$ ($L$ the linear lattice size) and $\Delta \tau=\sqrt{4/|U|}/10$. The convergences with respect to $\Theta$, $\Delta\tau$ and trial wave functions $\ket{\Psi_T}$ are carefully checked before our practical simulations. {See Appendix \ref {appendix:calcultion_parameters} for details.}

In DQMC, it is necessary to transform the interaction terms into the coupling between the fermion density and auxiliary bosons through a suitable HS decomposition. For our purpose, we rewrite the $U$- and $V$-terms as
\begin{equation} \label{eq:newHS}
	H_I=\frac{g}{2}\sum_{\langle ij\rangle}[(n_i-1)+a(n_j-1)]^2,
\end{equation}
{where $g$ and $a$ are related to $U$ and $V$ through the relations $U=\frac{gz}{2}(1+a^2)$ and $V=ga$ with $z$ the coordinate number. The solution of $a$ exists as long as $\frac{|U|}{|V|}=\frac{z}{2}|\frac{1}{a}+a|\ge z$. For the square lattice under concern, $z=4$, corresponding to $|U|\ge 4|V|$.} Eq.~\ref{eq:newHS} can be used to perform HS decoupling directly. For a particular bond $\langle ij\rangle$, let us denote $A=n_i-1+a (n_j-1)$. Then the HS transformation is performed as \cite{dqmc},
\begin{equation} \label{eq:HSgeneral}
	\text{e}^{-\frac{\Delta\tau g}{2} A^2}=\sum_{s=\pm1,\pm2}\gamma_{s}\text{e}^{\lambda_{s}A}+\mathcal{O}(\Delta \tau ^4),
\end{equation}
where $\gamma_{\pm1}=\frac14+\frac{\sqrt{6}}{12}$, $\gamma_{\pm2}=\frac14-\frac{\sqrt{6}}{12}$, $\lambda_{\pm1}=\pm\sqrt{-\Delta\tau g(3-\sqrt{6})}$, and $\lambda_{\pm2}=\pm\sqrt{-\Delta\tau g(3+\sqrt{6})}$. If $g<0$, $\lambda_s$ is real, the up and down determinants are real and the same, leading to the absence of sign problem. If $g>0$, $\lambda_s$ becomes purely imaginary, but {on a PH symmetric bipartite lattice,} the partial PH transformation can be used to prove the absence of the sign problem. Therefore, there is no sign problem in all our HS decoupling region $|U|\ge 4|V|$. {See Appendix \ref{appendix:sign} for details about the sign problem.} The accessible parameter space is the same as in Ref.~\cite{2d6}, but our HS fields live on bonds alone, while both on-site and on-bond fields are introduced in Ref.~\cite{2d6}.

The above HS decoupling enables us to perform the DQMC most efficiently. In order to search for possible ordering tendencies, we calculate the structure factor
\begin{equation}
S_{\hat{O}}(\bm{q})=\frac{1}{L^4}\sum_{i,j}\text{e}^{\text{i}\bm{q}\cdot(\bm{r_i-r_j})} \left( \langle \hat{O}_i\hat{O}_j^{\dagger}\rangle - \langle \hat{O}_i\rangle\langle\hat{O}_j^{\dagger}\rangle\right),
\end{equation}
where $L$ is the linear lattice size, $\hat{O}_i$ is a fermion bilinear operator in a given channel, and ${\bm q}$ is the ordering momentum. Note that according to the above normalization, a finite value of $S_{\hat{O}}$ in the thermodynamic limit implies the corresponding long range order (at the associated momentum), whereas it vanishes if such an order is not present. For AFM, $\hat{O}_i$ is the local spin operator $\bm{S}_i=\frac{1}{2}c_i^{\dagger}\bm{\sigma}c_i$ and $\bm{q}=\bm{Q}=(\pi, \pi)$; For CDW, $\hat{O}_i=n_i-1$ and $\bm{q}=\bm{Q}$; For s-wave SC (s-SC), $\hat{O}_i = c_{i\downarrow}c_{i\uparrow}$ and $\bm{q}=0$. For d-wave SC (d-SC), $\hat{O}_i= \frac{1}{\sqrt{2}} (c_{i\downarrow} c_{i+\mathbf{x}\uparrow} -c_{i\downarrow} c_{i+\mathbf{y}\uparrow})$ and $\bm{q}=\bm{0}$. For PS, $\hat{O}_i=n_i-1$. Strictly speaking, PS is signaled by the charge structure factor at zero momentum, {which is just the fluctuation of the total particle number $N$ since $\sum_{ij} (\av{n_in_j}-\av{n_i}\av{n_j}) =\av{N^2}-\av{N}^2$, hence, is exactly zero in the canonical ensemble}. However, PS can be probed by the charge structure factor at a small momentum $\delta \bm{q}$ in the finite-size system. We tested two choices,  $\delta\bm{q}=\bm{q}_x=(\frac{2\pi}{L},0)$, or $\delta\bm{q}=\delta\bm{q}_{xy}=(\frac{2\pi}{L},\frac{2\pi}{L})$, and we find that consistently, the structure factor $S_{\rm PS}(\delta\bm{q}_x)$ is always larger than $S_{\rm PS}(\delta\bm{q}_{xy})$ near or inside the PS phase. Therefore, in the following, only the PS structure factor $S_{\rm PS}(\delta\bm{q}_x)$ will be presented.
{In practice, we choose lattice size $L=4,6,8,10,12,16$ to calculate various structure factors and then perform finite size extrapolation to obtain the results in the thermodynamic limit.
If any one structure factor keeps finite after extrapolation, the corresponding long range order is formed by breaking relevant symmetries: AFM breaks spin rotation and translation, CDW and PS break translation, s-SC and d-SC break U(1) gauge symmetry.}

We also study the dynamical properties. The single particle gap $\Delta_{sp}$ can be extracted from the Matsubara Green's function
$G_{\sigma }(\bm{k},\tau)=-\langle T_{\tau} c_{\sigma}(\bm{k},\tau)c^{\dagger}_{\sigma}(\bm{k},0)\rangle$ ($T_\tau$ the time ordering operator)
through the long time behavior $G_{\sigma }(\bm{k},\tau)\propto\exp(-\tau \Delta_{sp}(\bm{k}))$. Similarly, the spin gap can be extracted from the Matsubara spin-spin correlation function
$\chi_s(\bm{q},\tau)= -\langle T_{\tau}  \bm{S}(\bm{q},\tau)\cdot \bm{S}(\bm{-q},0)\rangle$.
To get more information about spin excitations, we further adopt the maximum entropy method \cite{ana} to perform analytic continuation to extract the dynamic spin susceptibility $\chi_{s}(\bm{q},\omega)$.
{For the dynamical calculations, we choose lattice $L\le12$.}


\section{RESULTS AND DISCUSSION}\label{section3}

\begin{figure}
	\centering
	\includegraphics[width=0.48\textwidth]{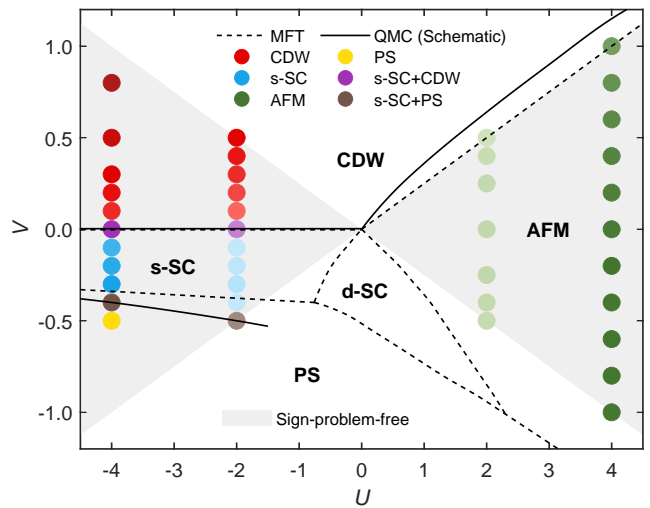}
	\caption{The ground state phase diagram of 2D half filled extended Hubbard model on the square lattice, as a function of on-site $U$ and nearest-neighbor interaction $V$. The color-coded circles represent different structure factors $S(\bm{q})$ by extrapolating the DQMC results to the thermodynamic limit $L\to\infty$. The regions  free of sign problem are indicated by shaded areas. The dashed lines are phase boundaries in the MFT and the solid lines are inferred from the DQMC. {For $U<0$, the DQMC (solid line) and MFT (dashed line) phase boundaries between s-SC and CDW are both the line of $V=0$ but slightly separated for clarity.}}
	\label{phase}
\end{figure}

Our main results are summarized in the phase diagram shown in Fig.~\ref{phase}. The shaded area indicates the accessible region free of sign problem in our DQMC. The colored circles show the DQMC results after extrapolation to the thermodynamic limit $L \rightarrow \infty$. Different orders are colored differently, and the color intensity corresponds to the value of the structure factor $S(\bm{q})$ normalized by the maximal value. The solid lines are the phase boundaries, which are obtained by DQMC in the accessible region, and by arguments to be clearer when they are out of the accessible region. The dashed lines are phase boundaries from MFT for comparison.

\begin{figure}
	\centering
	\includegraphics[width=0.48\textwidth]{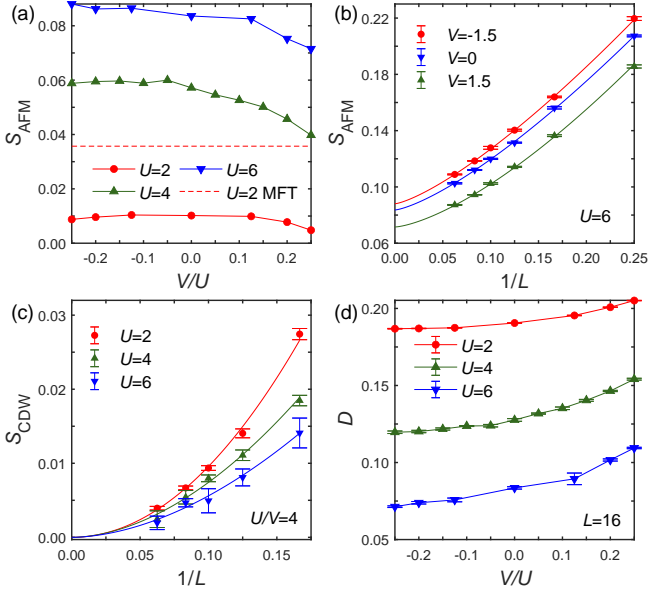}
	\caption{DQMC results for $U>0$. (a) $S_{\rm AFM}$ in the thermodynamic limit versus $V/U$ by fixing $U=2$, $4$ and $6$, respectively. The dashed line denotes the same quantity from MFT. (b) Finite size extrapolation of $S_{\rm AFM}$ versus $1/L$ for fixed $U$ and $V$. (c) Finite size extrapolation of $S_{\rm CDW}$ at $U=4V$. (d) Double occupancy $D$ at $L=16$ as a function of $V/U$ by fixing $U=2$, $4$ and $6$, respectively.}
	\label{sdw}
\end{figure}

\subsection{AFM phase for $U>0$}
Let us first look at the AFM phase in the case of positive $U$. In Fig.~\ref{phase}, the green circles indicate the AFM structure factor $S_{\rm AFM}$, which increases with $U$ and decreases as $V$ approaches $U/4$. The values of $S_{\rm AFM}$ in the thermodynamic limit are shown in Fig.~\ref{sdw}\blue{(a)}, where the symbols are DQMC results for $U=2,4,6$, and the dashed line represents the MFT result for $U=2$. Clearly, the MFT cannot capture the effect of $V$ on AFM ordering as it does not even vary with $V$.
Instead, in the DQMC results, we find that when $V>0$, $S_{\rm AFM}$ decreases with $V$ for a given $U$. When $V$ reaches the MFT phase boundary $U/4$, $S_{\rm AFM}$ remains finite, as also shown in Fig.~\ref{sdw}\blue{(b)}. This means AFM survives at $U=4V>0$. To see whether the CDW enters here, we show $S_{\rm CDW}$ in Fig.~\ref{sdw}\blue{(c)}. It clearly extrapolates to zero in the thermodynamic limit. Therefore, our DQMC results imply that the AFM-CDW transition must occur at $V>U/4$ (although it is beyond our sign-free space), in agreement with the cluster-based simulations \cite{vcpt, dca1, dca2, dca3,dba}. This boundary is shown schematically as a solid line in the first quadrant in Fig.~\ref{phase}.
On the other hand, from Fig. \ref{sdw}\blue{(a)}, we see that  $S_{\rm AFM}$ changes only mildly versus negative $V$, and is overall larger than the value at positive $V$. This tendency is more obvious for larger values of $U$. To see why this is the case, we examine the double occupancy $D=\frac{1}{N}\sum_{i}\langle n_{i\uparrow}n_{i\downarrow}\rangle$, as shown in Fig.~\ref{sdw}\blue{(d)}.
As $V$ decreases, the double occupation $D$ is reduced, and this is equivalent to the tendency toward development of local moment, which orders in the AFM phase. This tendency is eventually overwhelmed by a sufficiently large and negative $V$, upon which the PS or d-SC phase sets in, according to the MFT. Unfortunately this region is beyond our sign-free zone. We have checked the absence of d-SC in our QMC region, as shown in Appendix \ref{appendix:dSC}.
We also note that for $U=2$, $S_{\rm AFM}$ decreases slightly for $V<-0.15$. This behavior cannot be explained by the formation of local moment, and indeed, at this level the double occupancy is close to the free value $1/4$, and is better described in terms of itinerant antiferromagnetism \cite{FazekasBook,FradkinBook}.

\begin{figure}
	\centering
	\includegraphics[width=0.48\textwidth]{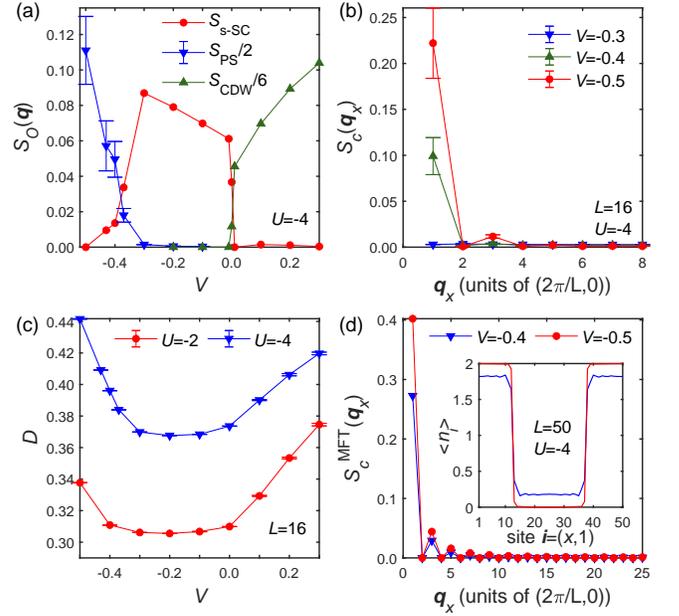}
	\caption{DQMC results for $U<0$. (a) $S_{\rm s-SC}$, $S_{\rm CDW}/6$ in the thermodynamic limit and $S_{\rm PS}/2$ at $L=16$ as functions of $V$ with $U=-4$. (b) Dependence of $S_c(\bm{q}_x)$ on $\bm{q}_x$ for $V=-0.3$, $-0.4$ and $-0.5$, respectively. (c) Double occupancy $D$ as a function of $V$ with fixed $U=-2 $ and $-4$ at $L=16$ . (d) $S_c^{\text{MFT}}(\bm{q}_x)$ from MFT at $U=-4.0$ and $V=-0.4$, $-0.5$. The inset shows a typical PS configuration by plotting the particle number $\langle n_{i} \rangle $ along a line cut $(i,1)$.
	}
	\label{u_negetive}
\end{figure}

\subsection{CDW, s-SC, and PS phases for $U<0$}
On the negative $U$ side, the transition between s-SC and CDW occurs exactly at $V=0$, as shown in Fig.~\ref{u_negetive}\blue(a) for $U=-4$.
{More results for $U=-2$ can be found in Appendix \ref{appendix:u_2}.} At $V=0$, the s-SC and CDW coexist and satisfy the relation $S_{\rm CDW}=2S_{\rm s-SC}$ exactly (not shown) as a result of the SO(4) symmetry of the negative-$U$ Hubbard model \cite{Yang1989,ZSC}.
In particular, the pseudo-SU(2) symmetry rotates the triplet of two branches of s-SC (complex) and one CDW (real), so the degeneracy between s-SC and CDW here is protected by symmetry. But this symmetry is immediately broken by $V$, which favors CDW (s-SC) if $V>0$ ($V<0$). The transition is between two ordered states, and is clearly first-order.
As shown in Fig.~\ref{u_negetive}\blue{(a)}, the structure factor $S_{\rm CDW}$ quickly grows up and $S_{\rm s-SC}$ suddenly vanishes as $V$ increases across zero. Their coexistence only occurs at $V=0$, which is a critical line for this first order transition.
On the other hand, as $V$ decreases further, the s-SC finally vanishes and is replaced by the PS phase, as shown in Fig.~\ref{u_negetive}\blue{(a)}.
For $V=-0.37, -0.40, -0.43$, s-SC and PS are found to coexist, indicating this phase transition is also first-order but occurs in a finite region. Note their coexistence is not protected by symmetries like the $V=0$ case.

Let us have a closer look into the PS state.
In Fig.~\ref{u_negetive}\blue{(b)}, we plot the density-density structure factor $S_c(\bm{q}_x)$ along a line cut $\bm{q}_x=(i\frac{2\pi}{L},0)$ with $i=0,1,\cdots,\frac{L}{2}-1$. When the system does not exhibit PS at $V=-0.3$, $S_c(\bm{q}_x)$ for all $\bm{q}_x$ are very small. While for $V=-0.4$ and $V=-0.5$, $S_c(\bm{q}_x)$ exhibits a peak at $\bm{q}_x=\delta\bm{q}_x$, indicating the tendency toward macro-scale density wave order, namely phase separation. (We did not perform finite-size scaling but use the largest size $L=16$, as near or inside the PS phase it is difficult to achieve nice ergodicity. {See Appendix \ref{appendix:PS} for more details}.) For comparison, the MFT result for PS are also shown in Fig.~\ref{u_negetive}\blue{(d)}. It appears to be similar to the DQMC result. The inset shows the mean particle number $\langle n_{i} \rangle $, which is constant in the y-direction. In the x-direction, half of the system is almost doubly occupied and the other half is empty.
In Fig.~\ref{u_negetive}\blue{(c)}, we present our DQMC result of double occupancy $D$, which is found to increase rapidly when $V=-0.4$ for $U=-4$, and $V=-0.5$ for $U=-2$. Combining $S_{\rm PS}$ and $D$, we conclude the system undergoes the PS transition between $V=-0.3$ and $V=-0.4$ in the case of $U=-4$.

\begin{figure}
	\centering
	\includegraphics[width=0.48\textwidth]{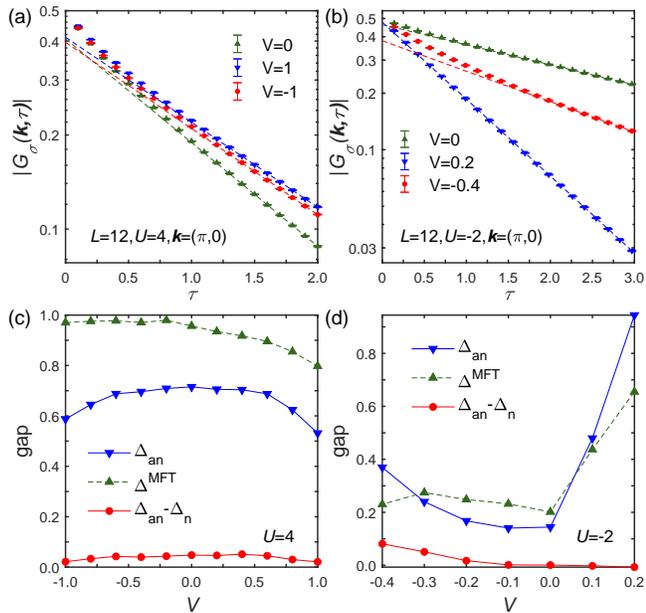}
	\caption{{Matsubara single particle Green's function $G_\sigma(\0k,\tau)$ for $U=4$ (a) and $U=-2$ (b), which are obtained at $L=12$. The single particle gap $\Delta_{\text{n,an}}$ can be extracted from the long time tail of $G_\sigma(\0k,\tau)\propto\text{exp}\left(-\Delta_{\text{n,an}}\tau\right)$ fitted as dashed lines. The thermodynamic limit values of $\Delta_{\text{n,an}}$ are plotted in (c) for $U=4$ and (d) for $U=-2$, respectively.} The mean field single particle gap $\Delta^{\text{MFT}}$ is also presented for comparison.}
	\label{dynamic}
\end{figure}

\subsection{Dynamical excitations}
The Matsubara Green's function $G_\sigma(\bm{k},\tau)$ is presented in Figs.~\ref{dynamic}\blue{(a,b)}. The single particle gap $\Delta_{sp}$ can be extracted from the long time behavior $G_\sigma(\bm{k},\tau)\sim \text{exp}\left[-\tau\Delta_{sp}(\bm{k})\right]$. We mainly focus on $\Delta_{\text{an}}$ at the anti-nodal point $(\pi,0)$, and $\Delta_{\text{n}}$ at the nodal point $(\frac{\pi}{2},\frac{\pi}{2})$.
For comparison, we take $\sqrt{S_{\hat{O}}}$ obtained from DQMC as the order parameter, and substitute it into the MFT Hamiltonian to compute the single particle gap $\Delta^{\text{MFT}}$, which is the same for nodal and antinodal points for the phases accessible by our DQMC.

For the AFM state at $U=4$, the single particle gap $\Delta$ as a function of $V$ is shown in Fig.~\ref{dynamic}\blue{(c)}. Although in MFT, $\Delta^{\text{MFT}}$ is proportional to the order parameter, which decreases monotonically with $V$, the DQMC gap $\Delta_{\text{an}}$ shows a non-monotonic dependence. We also observe a mild gap anisotropy $\Delta_{\text{an}}-\Delta_{\text{n}}$ (red dots).
These features indicate that the exact AFM ground state has an internal structure beyond the determinant state in MFT.

The results of negative $U$ with $U=-2$ are shown in Fig. \ref{dynamic}\blue{(d)}. A positive $V$ induces CDW, and the single particle gap $\Delta$ increases with $V$ for both DQMC and MFT, but the DQMC gap grows faster than mean field gap, indicating the correlation effect dominates for larger $V$.
For negative $V$, the system enters the s-SC and PS states successively as $V$ decreases. We observe that $\Delta_{\text{n,an}}$ increases more slowly than $\Delta^{\text{MFT}}$ for small $|V|$, but much faster for large $|V|$ as the PS phase is approached.
It is also interesting to compare the nodal and anti-nodal gaps: $\Delta_{\text{an}}$ is found to be always larger than $\Delta_{\text{n}}$ in the AFM and s-SC states, but not in the CDW state.

\begin{figure}
	\centering
	\includegraphics[width=0.48\textwidth]{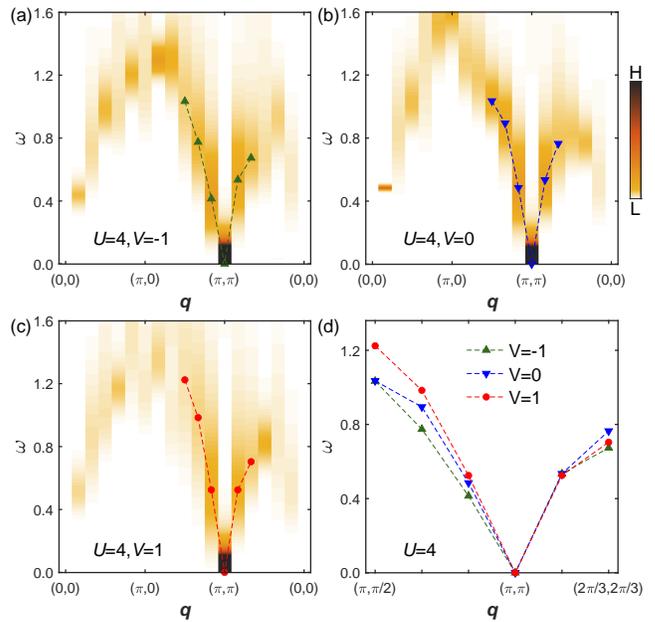}
	\caption{Spin excitation spectra along a high-symmetry path in the Brillouin zone for $U=4$ and $L=12$, and $V=-1$ (a), $V=0$ (b), or $V=1$ (c). The filled symbols represent the local maxima in $\bm{q}$ near $(\pi,\pi)$, defining the spin-wave dispersion, collected and shown in (d) for comparison.}
	\label{ss}
\end{figure}

{Finally, we perform analytic continuation of the Matsubara spin-spin correlation function $\chi_s(\0q,\tau)$ with the maximum entropy method \cite{ana} to obtain the spin excitation spectra as shown in Fig.~\ref{ss}.} The excitation energy goes to zero linearly near the ordering momentum $(\pi,\pi)$, a clear indication of the Goldstone mode. The spectral function is significantly smeared up near the zone boundary, indicating nontrivial magnon scattering effect (in the local moment limit) or multiple particle-hole excitation (in the itinerant limit). We also show the dispersion of the spin wave in Fig.~\ref{ss}\blue{(d)}, which is obtained from the local peaks (symbols) in Figs.~\ref{ss}\blue{(a-c)}. Clearly the slope, or the spin-wave velocity, is largest for $V=1$. {This feature seems to be associated with the spin-exchange interaction $t^2/(U-V)$ anticipated in the Mott limit.}


\section{CONCLUSION}\label{section4}
We presented a form of HS decomposition to deal with the extended Hubbard model with DQMC. On this basis,
we systematically investigated the ground state properties of the half-filled extended Hubbard model on the square lattice by zero-temperature DQMC simulations. We obtained the complete phase diagram in the sign-free parameter space $|U|\ge4|V|$, and discussed the dynamical excitations in some particular phases. Phase boundary beyond the sign-free space is also argued. These results should be important to benchmark further studies using other approaches.

We remark that our HS transformation can be generalized to longer range Coulomb interactions easily. For example, if the next-nearest-neighbor Coulomb interaction $V'$-term is added, we can rewrite the interactions including $U$, $V$ and $V'$ as
\begin{align} \label{eq:plaquette}
H_I=-\frac{g}{2}\sum_{i}[&Q_{i}+aQ_{i+\mathbf{x}}+aQ_{i+\mathbf{y}}+a^2Q_{i+\mathbf{x}+\mathbf{y}}]^2,
\end{align}
where $Q_i=n_i-1$, and $\mathbf{x},\mathbf{y}$ are primitive translation vectors. We can solve $(g,a)$ in terms of $(U,V,V')$. The subsequent HS decoupling is then straightforward, yielding HS fields living on centers of plaquettes. Work in this direction is in progress.

\section{ACKNOWLEDGMENTS}
M. Y. thanks Rui-Ying Mao and Qing-Geng Yang for helpful discussions. This work is supported by National Natural Science Foundation of China (under Grants Nos. 11874205, 12274205, and 11574134). The numerical simulations were performed in the High-Performance Computing Center of Collaborative Innovation Center of Advanced Microstructures, Nanjing University.

\appendix
\section{MFT calculations}
\label{appendix:MFT}

For the calculations of MFT, we mainly follow Ref.~\cite{Dagotto} to determine the ground states of the extended Hubbard model and to obtain the magnitude of each order parameter.
We choose the MFT ansatz for different orders as
\begin{align}
\text{SDW}:&\quad \frac12\av{n_{i\up}-n_{i\dn}}=(-1)^i m, \\
\text{CDW}:&\quad \av{n_{i\up}+n_{i\dn}-1} =(-1)^i Q, \\
\text{PS}:&\quad
\av{n_{i\up}+n_{i\dn}-1} =\sum_\0q\me^{-i\0q\cdot\0r_i}P_\0q, \\
\text{s-SC}:&\quad
\av{c_{i\dn}c_{i\up}} =\Delta_s, \\
\text{d-SC}:&\quad
\av{c_{i\dn}c_{i\pm\0x\up}}=\Delta_d, \,\, \av{c_{i\dn}c_{i\pm\0y\up}}=-\Delta_d,
\end{align}
where $m$, $Q$, $P_\0q$, $\Delta_s$, $\Delta_d$ are order parameters of AFM, CDW, PS, s-SC, d-SC, respectively.
Note that for PS, we consider a series of $\0q=(2i\pi/L,0)$ with $i=1,2,\cdots$.
After choosing one of the MFT ansatz, we can decouple the interaction terms in the relevant channel to obtain a mean field bilinear Hamiltonian, from which the order parameter can be solved self-consistently. For a given group of $(U,V)$, the ground state is determined by comparing the energies of different ansatz. Finally, by scanning the $(U,V)$ plane with a step length $0.1t$, we obtain the MFT phase diagram shown in Fig.~\ref{phase} in the main text. In our MFT calculations, we set the lattice size $L=100$.


\section{Sign problem}
\label{appendix:sign}

Our HS decoupling scheme can be applied to both finite and zero temperatures. In this work, since we are interested in the ground state properties, we employ the projector version at zero temperature. We first introduce the algorithm and conditions for absence of the sign problem in the extended Hubbard model. Then, we show the sign average by comparing our HS decoupling method with two other schemes in Refs.~\cite{2d6,2d1} inside or outside of our sign problem free region.

\subsection{projector DQMC algorithm}
The expectation value of an operator $\hat{O}$ in the ground state $\ket{\Psi_0}$, as shown in Eq.~\ref{eq:expection}, can be expressed as \cite{dqmc}
\begin{equation}\label{eq:p}
	\frac{\bra{\Psi_T}\text{e}^{-\Theta H/2} \hat{O} \text{e}^{-\Theta H/2}\ket{\Psi_T}}{\bra{\Psi_T}\text{e}^{-\Theta H}\ket{\Psi_T}}=\frac{1}{Z}\sum_{\bm{s}}\underbrace{A_{\bm{s}}D_{\bm{s}}^{\up} D_{\bm{s}}^{\dn}}_{p_{\bm{s}}}\langle \hat{O}\rangle_{\bm{s}} ,
\end{equation}
where $Z=\sum_{\bm{s}}p_{\bm{s}}=\sum_{\bm{s}}A_{\bm{s}}D_{\bm{s}}^{\up} D_{\bm{s}}^{\dn}$. The symbol $\bm{s}$ denotes a set of HS fields $\{s_{\langle ij \rangle,m}\}$ on each nearest-neighbor bond $\langle ij \rangle$ and time slice $m$, as introduced by the HS transformation of Eq.~\ref{eq:HSgeneral}.
The prefactor $A_{\bm{s}}$ is
\begin{equation}\label{eq:A}
	A_{\bm{s}}=\prod_{\langle ij \rangle,m}\gamma_{\langle ij \rangle,m}\exp[-{\lambda}_{\langle ij \rangle,m}(1+a)] .
\end{equation}
In Eq.~\ref{eq:p}, $D_{\bm{s}}^\sigma$ are determinants for spin $\sigma=\up,\dn$, respectively, given by
\begin{align}\label{eq:D}
	D_{\bm{s}}^\sigma=\det(P^{\sigma,\dagger}B_{M+1}^{\sigma}B_{M}^{\sigma}B_{M-1}^{\sigma}\cdots B_{1}^{\sigma}B_{0}^{\sigma}P^{\sigma}),
\end{align}
with
\begin{align}
	B_0^{\sigma}&=\exp \left(-\frac{1}{2}\Delta \tau K^{\sigma}\right),\,
	B_{M+1}^{\sigma}=\exp \left(\frac{1}{2}\Delta \tau K^{\sigma}\right), \\
	B_m^{\sigma}&=\exp \left(-\Delta \tau K^{\sigma}\right)\exp (V_m^{\sigma}),\quad m=1,2,\cdots,M ,
\end{align}
where $K^{\sigma}$ is the kinetic matrix defined as $\sum_{i,j;\sigma}c_{i \sigma}^{\dagger}K_{ij}^{\sigma}c_{j \sigma}$ and $V_m^{\sigma}$ is the potential matrix under a given HS field configuration defined as
\begin{equation} \label{eq:hs}
	\sum_{\langle ij \rangle} {\lambda}_{\langle ij \rangle,m}(\hat{n}_i+a\hat{n}_j)=\sum_{i,\sigma}c_{i\sigma}^\dag(V_{m}^{\sigma})_{ii}c_{i \sigma} .
\end{equation}
The matrices $P^{\sigma}$ in Eq.~\ref{eq:D} is given by a trial Hamiltonian $H_T=\sum_{\av{ij}\sigma}c_{i\sigma}^\dag (h_T^\sigma)_{ij} c_{j\sigma}$.
That is, $P_{:,n}^\sigma$ is the lowest $n$-th eigenstate of $h_T^\sigma$, with $n\le N_p$ ($N_p$ the particle number). Note that $P^\sigma$ is the same for up and down electrons since we work in total spin $S_z=0$ space.
In practice, we simply choose $h_T^\sigma=K^\sigma$ but add
additional small random hoppings on each nearest neighbor bond to eliminate possible energy level degeneracies.
Since $h_T^\sigma$ is purely real and spin-independent, we can construct real $P^\dn=P^\up$.
%

\subsection{Sign problem free at $g<0$}
Next, let us investigate the sign problem.
When the interaction $g<0$, the HS field ${\lambda}_{\langle ij \rangle,m}$ defined in Eq.~\ref{eq:HSgeneral} is purely real. In this case, we have
\begin{align}
K^\dn=K^\up,\quad V_{\bm{s}}^\dn=V_{\bm{s}}^{\up},\quad P^\dn=P^\up,
\end{align}
from which we obtain
\begin{align}
D_{\bm{s}}^\dn=D_{\bm{s}}^\up,
\end{align}
which is also real.
In addition, we also have
\begin{align}
A_{\bm{s}}>0,
\end{align}
from its definition Eq.~\ref{eq:A}. Therefore, for any HS configuration ${\bm{s}}$, we have now proved
\begin{align}
p_{\bm{s}}=A_{\bm{s}}D_{\bm{s}}^{\up}D_{\bm{s}}^{\dn}=A_{\bm{s}}\left( D_{\bm{s}}^{\up}\right)^2\ge0 ,
\end{align}
since $D_{\bm{s}}^\up$ is purely real.

\subsection{Sign problem free at $g>0$ with particle-hole symmetry}
When the interaction $g>0$, ${\lambda}_{\langle ij \rangle,m}$ defined in Eq.~\ref{eq:HSgeneral} becomes purely imaginary.
If there is a PH symmetry, we can perform a partial-PH transformation for spin down electrons
\begin{align}
c_{i\dn}\to(-1)^i\tilde{c}_{i\dn}^\dag ,
\end{align}
after which we obtain
\begin{align}
\tilde{K}^\dn=K^\up,\quad \tilde{V}_{\bm{s}}^\dn=V_{\bm{s}}^{\up,*},\quad \tilde{P}^\dn=P^\up ,
\end{align}
where we add tilde to label matrices after the transformation.
From the above, we obtain
\begin{align}
\tilde{D}_{\bm{s}}^\up=D_{\bm{s}}^\up,\quad
\tilde{D}_{\bm{s}}^\dn=D_{\bm{s}}^{\up,*} .
\end{align}
In addition, the prefactor $A_{\bm{s}}$ should be transformed to
\begin{align}
\tilde{A}_{\bm{s}}=\prod_{\langle ij \rangle,m}\gamma_{\langle ij \rangle,m}>0 .
\end{align}
Therefore, we have now proved
\begin{align}
p_{\bm{s}}=\tilde{A}_{\bm{s}}\tilde{D}_{\bm{s}}^{\up}\tilde{D}_{\bm{s}}^{\dn}=\tilde{A}_{\bm{s}}D_{\bm{s}}^{\up}D_{\bm{s}}^{\up,*}\ge0 .
\end{align}

\begin{figure}
	\centering
	\includegraphics[width=0.48\textwidth]{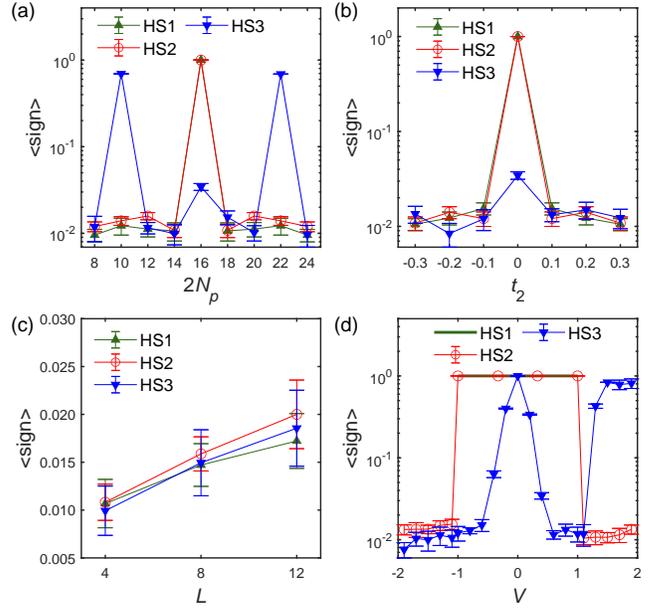}
	\caption{The average sign as functions of the particle number $2N_p$ (a), next-nearest neighbor hopping $t_2$ (b), lattice size $L$ (c) and density-density interaction $V$ (d), by comparing three HS decoupling schemes: HS1 (ours), HS2 (Ref.~\cite{2d6}), HS3 (Ref.~\cite{2d1}). We have fixed $U=4, \Theta=3L, \Delta \tau=0.1$. The other parameters are:  $L=4, V=0.4, t_2=0$ in (a); $L=4, V=0.4, 2N_p=16$ in (b); $V=0.4, 2N_p/N=7/8$, $t_2=0$ in (c); $L=4, 2N_p=16, t_2=0$ in (d). }
	\label{sign}
\end{figure}

\subsection{Sign problem at $g>0$ in general}
We have learned that when $g>0$, the sign problem can be free if there is a PH symmetry.
In this subsection, we examine the severity of the sign problem once the PH symmetry is broken.
We consider two ways to break the PH symmetry: away from half-filling or adding a next nearest neighbor hopping $t_2$.
We compare three schemes of HS decouplings. We use HS1 to represent our proposal (with only on-bond fields), HS2 for the method of Ref.~\cite{2d6} (with both on-site and on-bond fields), and HS3 for the early scheme of Ref.~\cite{2d1} (with real but spin-dependent fields).

We plot the sign average versus filling in Fig.~\ref{sign}(a) and versus $t_2$ in Fig.~\ref{sign}(b). It can be seen that as long as the PH symmetry is broken, all three schemes give very small sign average, unless for special fillings using HS3, making the DQMC simulation almost inaccessible. In Fig.~\ref{sign}(c), we check the size dependence, when sign problem occurs, which are all severe for the three methods.
Finally, we also examine the sign average with respect to $V$, as shown in Fig.~\ref{sign}(d).
When $|V|\le|U|/4$, HS1 and HS2 are both free of sign problem, but for HS3 the sign problem occurs immediately as $V\ne0$.
Comparing with HS2, our proposal (HS1) has fewer HS fields and thus easier to implement in practice.
When $|V|\ge|U|/4$, our scheme (HS1) does not apply (since no solution exists for the HS parameters), and HS2 has severe sign problem.
For HS3, the sign problem is still very severe for negative $V$ but seems to be weakened for relatively large $V$, making it a better candidate in that region.

\section{Benchmarks of our simulations}
\label{appendix:calcultion_parameters}
In this section, we provide benchmarks of the choices of discretization time $\Delta\tau$, projection time $\beta$, and trial wave function $\ket{\Psi_T}$.

\subsection{Discretization time $\Delta\tau$}
\begin{figure}
	\centering
	\includegraphics[width=0.48\textwidth]{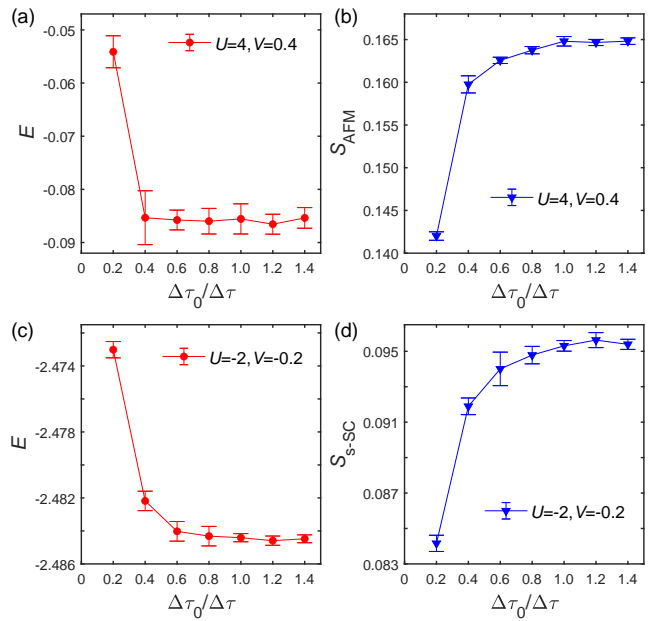}
	\caption{The ground state energy $E$ and AFM structure factor $S_{\rm AFM}$ are plotted with respect to $\Delta\tau_0/\Delta\tau$, with $\Delta\tau_0=\sqrt{4/U}/10$ as our practical choice in simulations. (a) and (b) are for $(U,V)=(4,0.4)$. (c) and (d) are for $(U,V)=(-2,-0.2)$. The lattice size is $L=4$.}
	\label{check_tao}
\end{figure}

In this work, we set $\Delta\tau$ to be independent of the lattice size \cite{white1989}, but depend on $U$ as $\Delta \tau=\sqrt{4/|U|}/10$. The convergences with respect to $\Delta\tau$ have been checked by comparing results with different $\Delta\tau$ in our simulations.
For $L=4$ with $(U,V)=(4,0.4)$ or $(-2,-0.2)$, the ground state energy $E$ and the AFM structure factor $S_{\rm AFM}$ or s-SC structure factor $S_{\rm s-SC}$ are plotted with respect to $\Delta\tau$ in Fig.~\ref{check_tao}. It can be seen that the results do converge with our practical choice labeled by $\Delta\tau_0$ in these figures.

\subsection{Projection time $\Theta$}
\begin{figure}
	\centering
	\includegraphics[width=0.48\textwidth]{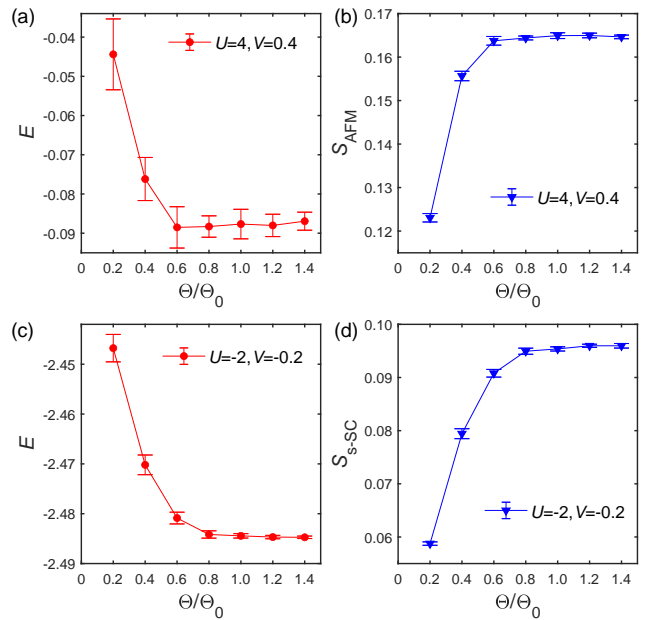}
	\caption{The ground state energy $E$ and AFM structure factor $S_{\rm AFM}$ are plotted with respect to $\Theta/\Theta_0$, with $\Theta_0=2L\sqrt{4/U}$ as our practical choice in simulations. (a) and (b) are for $(U,V)=(4,0.4)$. (c) and (d) are for $(U,V)=(-2,-0.2)$. The lattice size is $L=4$.}
	\label{check_theta}
\end{figure}

For projection time $\Theta$, we choose it proportional to the linear lattice size $L$ \cite{berg2012}, namely $\Theta=2L\sqrt{4/|U|}$. The convergences with respect to  $\Theta$ have been checked by comparing results with different $\Theta$ in our simulations. For $L=4$ with $(U,V)=(4,0.4)$ or $(-2,-0.2)$, we show the ground state energy $E$ and the AFM structure factor $S_{\rm AFM}$ or s-SC structure factor $S_{\rm s-SC}$ versus $\Theta$, as shown in Fig.~\ref{check_theta}. It can be seen that the results do converge with our practical choice labeled by $\Theta_0$ in these figures.

\begin{figure}
	\centering
	\includegraphics[width=0.48\textwidth]{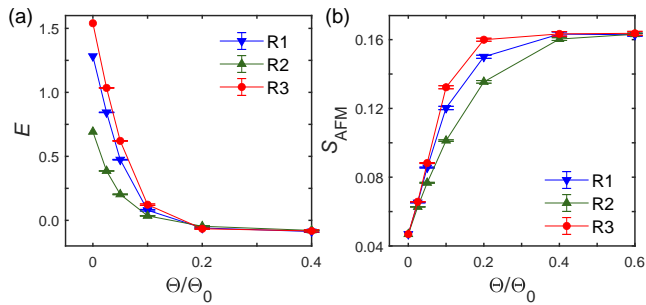}
	\caption{The ground state energy $E$ (a) and AFM structure factor $S_{\rm AFM}$ (b) as functions of $\Theta/\Theta_0$ with three randomly chosen trial wave functions R1, R2 and R3. The parameters are $L=4, U=4, V=0.4, \Delta \tau=0.1, \Theta_0=8$.}
	\label{different_trial}
\end{figure}

\subsection{Trial wave function $\ket{\Psi_T}$}
As for the trial wave function, we have selected several trial wave functions by adding different random hoppings on each nearest neighbor bond.
In Fig.~\ref{different_trial}, we plot the ground state energy $E$ and AFM structure factor $S_{\rm AFM}$ versus $\Theta$, respectively. The results are found to be independent of the trial wave functions for our practical choice $\Theta_0$.

\section{More QMC results}
In this section, we provide more QMC data, including the finite size effect of the PS structure factor, absence of d-wave SC, and some results of $U=-2$.

\begin{figure}
	\centering
	\includegraphics[width=0.48\textwidth]{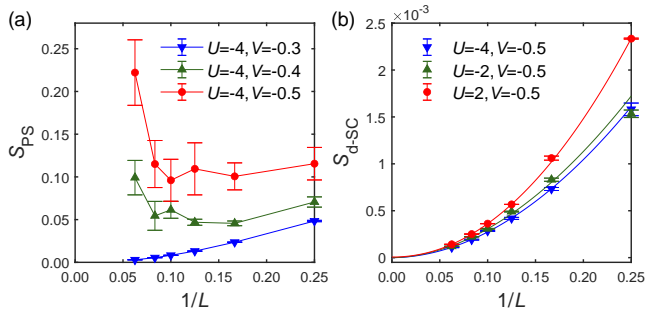}
	\caption{The structure factors $S_{\rm PS}$ (a) and $S_{\rm d-SC}$ (b) are plotted with respect to $1/L$.}
	\label{PS-dSC}
\end{figure}

\subsection{finite size effect of PS}
\label{appendix:PS}

In the simulations, we find it is difficult to achieve nice ergodicity near or inside the PS phase. The data quality is relatively poor and the error bars are relatively large, as shown in Fig.~\ref{PS-dSC}(a) for $U=-4$. When $V=-0.3$, the structure factor $S_{\rm PS}$ is quite smooth and extrapolates to zero, indicating the absence of the PS order. But for $V=-0.4$ and $V=-0.5$, the curves are qualitatively different from $V=-0.3$. $S_{\rm PS}$ even grows up as $L$ increases, indicating the formation of such a long range order. However, such a behavior causes difficulty to perform reliable finite size extrapolation. Therefore, we adopt a compromise way to use the data of $L=16$ to characterize the strength of the PS order, as discussed in the main text.
%

\subsection{Absence of d-wave SC}
\label{appendix:dSC}
Although there is a region with d-SC order in the MFT phase diagram, it is beyond our QMC accessible region without the sign problem. Nevertheless, we have examined the possibility of d-SC in our simulated region. In Fig.~\ref{PS-dSC}(b), we show the finite size extrapolation of $S_{\rm d-SC}$ versus $1/L$ for several groups of $(U,V)$ close to the MFT d-SC phase. All the curves extrapolate to zero, indicating the absence of d-SC in the thermodynamic limit inside our QMC accessible region.

\begin{figure}
	\centering
	\includegraphics[width=0.48\textwidth]{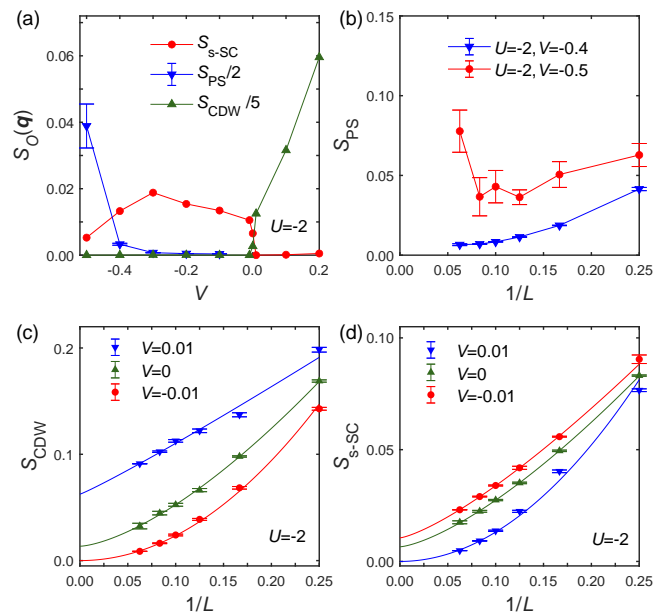}
	\caption{DQMC results for $U=-2$. (a) The structure factors of s-SC and CDW (after extrapolation), and PS ($L=16$) are plotted versus $V$. (b) shows $S_{\rm PS}$ versus $1/L$ for $V=0.4$ and $0.5$. (c) and (d) plot $S_{\rm CDW}$ and $S_{\rm s-SC}$, respectively, versus $1/L$ for $V=-0.01, 0, 0.01$. }
	\label{u_2}
\end{figure}

\subsection{results of $U=-2$}
\label{appendix:u_2}

In the main text, we have presented the data of negative $U$ with $U=-4$. By decreasing $V$, we obtain CDW, s-SC and PS, respectively. The first transition between CDW and s-SC occurs at $V=0$ exactly, while the second transition between s-SC and PS occurs at about $V=-0.4$ with a coexistent region.
Here we present the data for $U=-2$ directly, which confirm the results of $U=-4$ and have already been partially displayed in the phase diagram Fig.~\ref{phase}.

In Fig.~\ref{u_2}(a), we show the structure factors $S_{\rm s-SC}$, $S_{\rm CDW}$ (after finite size extrapolation), and $S_{\rm PS}$ (at L=16) with respect to $V$. Similar to the results of $U=-4$, we also find the two transitions. The transition between PS and s-SC occurs at $V<-0.4$, with a coexistent region. The finite size dependence of $S_{\rm PS}$ for $V=-0.5$ is quite different from $V=-0.4$, as shown in Fig.~\ref{u_2}(b), indicating the PS order is already established at $V=-0.5$.
For the transition between s-SC and CDW, we present their structure factors in Fig.~\ref{u_2}(c) and Fig.~\ref{u_2}(d), respectively. At $V=-0.01$ only $S_{\rm s-SC}\ne0$, and at $V=0.01$ only $S_{\rm CDW}\ne0$. They coexist only at $V=0$, strongly supporting the critical line at $V=0$ exactly as predicted by symmetry analysis.

\bibliography{reference.bib}
\end{document}